# Analysis of Extremely Obese Individuals Using Deep Learning Stacked Autoencoders and Genome-Wide Genetic Data

Casimiro A. Curbelo Montañez[1], Paul Fergus[1], Carl Chalmers[1] and Jade Hind[1]

[1] Liverpool John Moores University, Liverpool, UK
`{c.a.curbelomontanez,p.fergus, c.chalmers}@ljmu.ac.uk`
`j.hind@2012.ljmu.ac.uk`

**Abstract.** Genetic predisposition has been identified as one of the components contributing to the obesity epidemic in modern societies. The aetiology of polygenic obesity is multifactorial, which indicates that lifestyle and environmental factors may influence multiples genes to aggravate this disorder. Several low-risk single nucleotide polymorphisms (SNPs) have been associated with BMI. However, identified loci only explain a small proportion of the variation observed for this phenotype. The linear nature of genome wide association studies (GWAS) used to identify associations between genetic variants and the phenotype have had limited success in explaining the heritability variation of BMI and shown low predictive capacity in classification studies. GWAS ignores the epistatic interactions that less significant variants have on the phenotypic outcome. In this paper we utilise a novel deep learning-based methodology to reduce the high dimensional space in GWAS and find epistatic interactions between SNPs for classification purposes. SNPs were filtered based on the effects associations have with BMI. Since Bonferroni adjustment for multiple testing is highly conservative, an important proportion of SNPs involved in SNP-SNP interactions are ignored. Therefore, only SNPs with p-values $< 1\text{x}10^{-2}$ were considered for subsequent epistasis analysis using stacked auto encoders (SAE). This allows the nonlinearity present in SNP-SNP interactions to be discovered through progressively smaller hidden layer units and to initialise a multi-layer feedforward artificial neural network (ANN) classifier. The classifier is fine-tuned to classify extremely obese and non-obese individuals. The performance of classifications using progressively smaller compressed layers was compared and the results reported. The best results were obtained with 2000 compressed units (SE=0.949153, SP=0.933014, Gini=0.949936, Logloss=0.1956, AUC=0.97497 and MSE=0.054057). Using 50 compressed units it was possible to achieve (SE=0.785311, SP=0.799043, Gini=0.703566, Logloss=0.476864, AUC=0.85178 and MSE=0.156315).

# 1 Introduction

Obesity prevalence has increased over several decades and has now reached epidemic proportions [1]. This has had a significant impact on morbidity and mortality. Individuals suffering from obesity are at higher risk of developing numerous non-communicable diseases (NCDs), including type 2 diabetes, cardiovascular disease, and certain types of cancer [2].

The aetiology of common or polygenic obesity is multi-factorial, indicating that combinations of lifestyle and environmental factors may interact with multiple genes, causing this disorder. This is further supported by twin, adoption, and family studies which found that variation in body mass index (BMI) was largely due to heritable genetic differences, with heritability (the proportion of the variability of a trait that is attributable to genetic factors) estimates in adults ranging between 40% and 70% [3]. Hence, it is believed that obesity risk is higher among those individuals genetically predisposed to weight gain in obesogenic environments where gene-environment interactions occur.

The advent of high-throughput technologies has enabled hypothesis free approaches based on single-locus analysis such as genome-wide association studies (GWAS) [4]. In GWAS, Single Nucleotide Polymorphism (SNP) [5] are independently tested for association with a phenotype of interest, omitting, therefore, the existence of interactions between loci. Currently, several low-risk common genetic variants have been associated with BMI, including variants within the FTO or MC4R genes. However, these identified loci only explain a small proportion (~ 1.4-2.7%) of the variation observed in BMI [6]. The linear nature of the framework employed by GWAS may explain the limited success of these studies when explaining heritability variation of BMI. This is now regarded as a significant limitation in GWAS, particularly when studying complex disorders that rely on an understanding of gene-gene and gene-environment interactions [7]. An important challenge in the analysis of high-throughput genetic data is the development of computational and statistical methods to identify epistasis interactions. Investigating all combinations between SNPs in genome-wide studies is computationally very costly since the number of tests and time necessary to perform the exhaustive search increases exponentially with the order of interactions considered, making them not scalable. Therefore, epistatic analysis has primarily been restricted to two locust interactions.

In this paper a DL stacked auto encoder (SAE) is used to deal with nonlinearity present in SNP-SNP interactions and to initialise a multi-layer feedforward artificial neural network (ANN) classifier. The classifier is fine-tuned to classify extremely obese and non-obese individuals. Better approximations to non-linear functions can be generated by using DL models than those with a shallow structure. The combination of SAE and a multi-layer feedforward ANN is used to model the epistatic effects of a subset of filtered genetic variants (p-values $< 1 \times 10^{-2}$) based on their association with BMI after quality control (QC) procedures, demonstrating the potential of DL for GWAS.

The remainder of this paper is organized as follows. Section 2 describes the Materials and Methods used in the study. The results are presented in Section 3 and discussed in Section 4 before the paper is concluded and future work is presented in Section 5.

## 2 Materials and Methods

### 2.1 Study participants

Case and control data utilised in this study was requested from the database of Genotypes and Phenotypes (dbGaP) [8]. Participants were extracted from different study cohorts in the Geisinger MyCode project. A subset of 1,231 primary patients of a Geisinger Clinic with non-urgent visits to the clinic were used as control group. Conversely, 962 unique samples with a mean BMI of 49.17 (± 8.83 SD), part of a cohort of primary Caucasian patients from the Geisinger Clinic with extreme obesity who have undergone bariatric surgery were considered as control group.

A total of 2,193 participants of which 917 are males and 1,236 are females form the case/control dataset in this study. Each participant contains 594,034 markers. Furthermore, 99.5% of the participants belong to a white ethnical background (Caucasians).

### 2.2 Genetic analysis

**Quality Control**
To conduct association analyses, only those individuals reported to be white Caucasian were selected to reduce potential bias due to population stratification. Quality control, an imperative step prior to any GWAS analysis [9], was also applied to identify potentially problematic individuals and SNPs. The first step was to filter out and remove data samples with discordant sex. Related or duplicated samples were removed using Identity by Descent (IBD) coefficient estimates (IBD > 0.185). To obtain high-quality data for the association analysis, the data set was pruned with the following criteria: sample call rate >99%; SNP call rate >99%; and a threshold for Hardy–Weinberg equilibrium of 0.0001 in control cohort. Due to limited power of rare variants in an association study, we only kept SNPs with minor allele frequencies >0.05. After QC, 1,997 individuals (879 cases and 1,118 controls) and 240,950 genetic variants remained for subsequent analysis.

**Association Analysis**
Statistical association testing between individual SNPs and obesity was conducted under an additive model using logistic regression [10] in PLINK (v1.9) [11].

Let $i$ be the individuals ($i = 1, 2, \ldots, n$), $Y_i$ the phenotype for individual $i$ and $X_i$ the genotype of individual $i$ at a particular SNP. Let $Y \in \{0,1\}$ be a binary phenotype for case/control status and $X \in \{0,1,2\}$ be a genotype at the typed locus, where 0, 1 and 2 represent homozygous major allele $AA$, heterozygous allele $Aa$, and homozygous minor allele $aa$ respectively. Logistic regression modelling can be defined as a linear predictor function:

$$logit\big(E(Y_i \mid X_i)\big) \sim \beta_0 + \beta_1 X_i \qquad (1)$$

Correction for multiple testing is commonly adopted in association analysis, using adjustments such as Bonferroni correction [12]. However, it has been suggested that adjustments, such as Bonferroni, are too conservative and may result in missing significant associations when performing multiple association tests. Consequently, we did not adjust for multiple testing but rather considered the results of all association tests with p-values lower than $1 \times 10^{-2}$. Logistic regression is used therefore to reduce the number of SNPs to meet the computational needs required for epistatic analysis and machine learning tasks.

### 2.3 Multi-layer Feedforward Artificial Neural Network

A multi-layer feedforward ANN is implemented based on the formal definitions in [13], to conduct binary classification. Labelled training samples $(x^{(i)}, y^{(i)})$ from case-control genetic data are considered for a supervised learning problem. A complex non-linear hypothesis $h_{W,b}(x)$ is defined using the neural network, with parameters $W,b$ fitted to our data. The network first calculates the output matching the input (feed-forward stage). Then, the backpropagation algorithm [14] is used to calculate error propagating to previous layers, and finally, the weights of the network are adjusted.

Taking a set of labelled samples $\{x_1, x_2, ... x_n\}$ and a bias unit $b$ (+1 intercept term) as input, single computational units or neurons output

$$h_{W,b}(x) = f(W^T x) = f\left(\sum_{i=1}^{n} W_i x_i + b\right) \qquad (2)$$

where $f: \mathbb{R} \mapsto \mathbb{R}$ represents the activation function and $W$ the weight. Typically, a sigmoid function is used as an activation function although others such as hyperbolic tangent (tanh) or rectifier linear unit (ReLU) can be adopted.

Input, hidden and output layers make up the network structure where $l$ represent the total number of layers densely connected, $L_1$ the input layer and $L_{nl}$ the output layer. Several parameters constitute the neural network. The parameter $W_{ij}^{(l)}$ denotes the weight for the connection between the $j^{th}$ neuron in layer $l$, and the $i^{th}$ neuron in layer $l+1$. The bias unit $b_i^{(l)}$, associated with neuron $i$ in layer $l+1$, is introduced to counteract the problem associated with input patterns that are zero. The number of nodes in layer $l$ is denoted by $s_l$ without taking $b_i^{(l)}$ into consideration. Additionally, the activation or output value of node $i$ in layer $l$ of the network, denoted as $a_i^{(l)}$, is equal to the activation function of the total weighted sum of inputs (including the bias term), represented as $f(z_i^{(l)})$. Given that the values from the inputs are denoted by $a^{(1)} = x$ and the activation for layer $l$ is $a^{(l)}$, the activation output layer $l$ plus the intercept term +1 ($a^{(l+1)}$) can be computed. Thus, a compact vectorised form of $z_i^{(l)}$ and $a_i^{(l)}$ is given by

$$z^{(l+1)} = W^{(l)} a^{(l)} + b^{(l)} \qquad (3)$$

$$a^{(l+1)} = f(z^{(l+1)}) \qquad (4)$$

The neural network defines $h_{W,b}(x)$ which outputs a real number based on a given set of parameters $W,b$. Equations (3-4) can be used to compute the output of the network, successively calculating all the activations in layer $L_2$, then $L_3$ and so on up to the output layer $L_{nl}$.

Given a training set $\{(x^{(1)}, y^{(1)}),..., (x^{(m)}, y^{(m)})\}$ of $m$ samples, the neural network is trained using gradient descent optimisation and the overall cost function is defined as

$$J(W,b) = \left[\frac{1}{m}\sum_{i=1}^{m} J(W,b,x^{(i)},y^{(i)})\right] + \frac{\lambda}{2}\sum_{l=1}^{n_l-1}\sum_{i=1}^{s_l}\sum_{j=1}^{s_l+1}(W_{ji}^{(l)})^2$$
$$= \left[\frac{1}{m}\sum_{i=1}^{m}(\frac{1}{2}\|h_{W,b}(x^{(i)}) - y^{(i)}\|^2)\right] + \frac{\lambda}{2}\sum_{l=1}^{n_l-1}\sum_{i=1}^{s_l}\sum_{j=1}^{s_l+1}(W_{ji}^{(l)})^2 \quad (5)$$

where the first term is the average sum of squared errors and the second, a weight decay or regularization term that helps prevent overfitting by reducing the magnitude of the weights. The relative importance of the two expressions is controlled with the weight decay parameter $\lambda$. Next, backpropagation is used to efficiently compute the partial derivatives [13] of the cost function for a single sample with respect to any weight or bias in the network, $J(W,b;x,y)$. Finally, gradient descent is used to reduce our cost function $J(W,b)$ before training the neural network used in this study for classification purposes.

### 2.4 Autoencoders

Autoencoders (AE) belong to the unsupervised learning class of algorithms and can be used to pre-train neural networks [15]. A basic AE is a three-layer neural network that applies backpropagation to learn an output $\hat{x}$ that is similar to the input $x$. Hence, an AE tries to learn a function $h_{W,b}(x) \approx x$, given a set of unlabelled training samples $\{x^{(1)}, x^{(2)}, x^{(3)}, ...\}$, where $x^{(i)} \in \mathbb{R}^n$. An example of a single layer AE is illustrated in Fig.1, where the first and the third layers are the input and the reconstruction or output layer with 5 units, respectively. The second layer or hidden layer aims to generate the deep features by minimizing the error between the input vector and the reconstruction vector. Thus, an AE is a neural network with a single hidden layer composed by two parts, an encoder and a decoder.

The output of the encoder $h$ is a reduced representation of $x$ that is used by the decoder to reconstruct the original dataset $x$. The process is formulated as

$$h = a^{(2)} = f(W^{(1)}a^{(1)} + b^{(1)}), \quad (6)$$

$$x = a^{(3)} = f(W^{(2)}a^{(2)} + b^{(2)}) \quad (7)$$

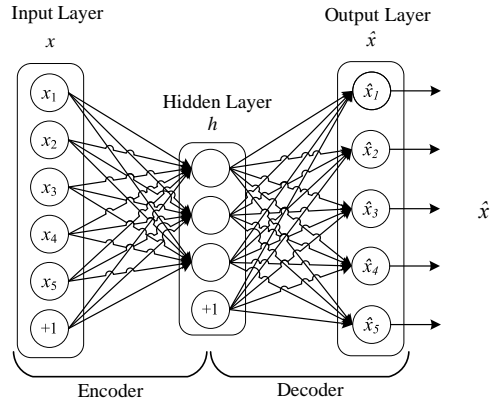

**Fig. 1.** Autoencoder diagram.

The AE will learn any structure present in the data. To improve the discovery of interesting patterns in the data in cases when the number of hidden neurons is elevated or even higher than the number of inputs, a sparsity constraint is imposed. This imposition introduces the concept of sparse autoencoders.

The sparsity constrain keeps the neurons inactive most of the time. The activation of the hidden neurons for a given input $x$ is denoted by $a_j^{(2)}(x)$. We let

$$\rho_j = \frac{1}{m} \sum_{i=m}^{m} \left[ a_j^2(x^{(i)}) \right] \tag{8}$$

represent the average activation of a hidden neuron ($j$) and enforce an approximation constraint $\hat{\rho_j}=\rho$, where $\rho$ is a small sparsity parameter typically close to zero. The activation of the hidden neurons requires to be mostly 0 to meet the aforementioned constrain. Hence, a penalty term that penalizes $\hat{\rho_j}$ is added, deviating significantly from $\rho$:

$$\sum_{j=1}^{s_2} \rho \log \frac{\rho}{\rho_j} + (1-\rho) \log \frac{1-\rho}{1-\rho_j}, \tag{9}$$

where the number of hidden neurons is defined by $s_2$, and $j$ is an index used to sum the hidden units in the network. Kullback-Leiber (KL) divergence between two Bernoulli random variables with mean $\rho$ and $\hat{\rho_j}$, is used to enforce the penalty term (9):

$$\sum_{j=1}^{s_2} \mathrm{KL}(\rho || \rho_j), \tag{10}$$

where

$$\mathrm{KL}(\rho || \rho_j) = \rho \log \frac{\rho}{\rho_j} + (1-\rho) \log \frac{1-\rho}{1-\rho_j} \tag{11}$$

Therefore, KL-divergence is used here as a function to measure the difference between two distributions, and it can be equal to 0 if $\hat{\rho_j}=\rho$ ($\mathrm{KL}(\rho||\hat{\rho_j})=0$) or it increases monotonically as $\hat{\rho_j}$ diverges from $\rho$. The overall cost function for the AE can be now defined as

$$J_{sparse}(W,b) = J(W,b) + \beta \sum_{j=1}^{s_2} \mathrm{KL}(\rho || \rho_j), \tag{12}$$

where $J(W,b)$ remains as defined earlier in this paper whilst $\beta$ is introduced to control the weight of the sparsity penalty term. Since the term $\hat{\rho_j}$ is the average activation of hidden neuron $j$, it is also dependent on $W,b$.

Finally, the KL-divergence term can be incorporated into the derivative calculation [13] by computing the error term:

$$\delta_i^{(2)} = \left( \sum_{j=1}^{s_3} W_{ji}^{(2)} \delta_j^{(3)} \right) f'(z_i^{(2)}) + \beta \left( -\frac{\rho}{\rho_i} + \frac{1-\rho}{1-\rho_i} \right), \tag{13}$$

Once a single layer AE has been trained, a second AE can be trained using the output layer of the first AE as shown in Fig. 2. By repeating this procedure, it is possible to create Stacked Autoencoders of arbitrary depth. The results produced by the SAE in this study are utilized to pre-train the weights for our multi-layer feedforward ANN (softmax model) to classify extreme cases of obesity and normal individuals. Quality control, association analysis, multi-layer feedforward ANN classifier and deep learning SAE constitute the components of our proposed methodology.

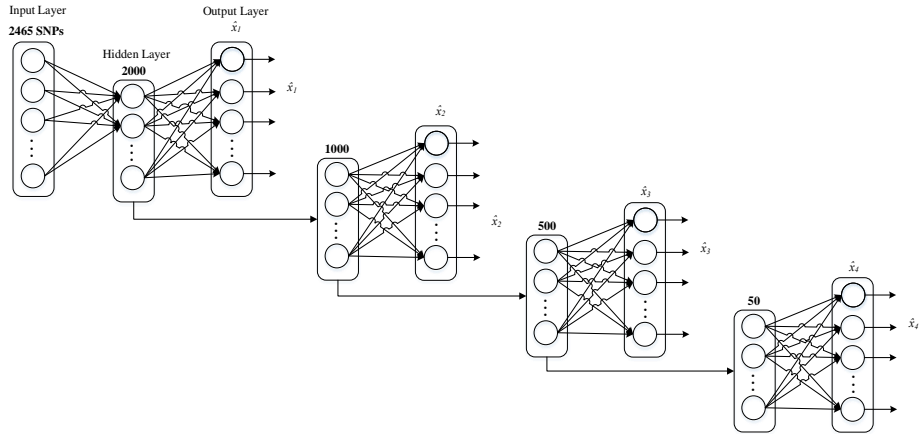

**Fig. 2.** Proposed SAE. Features are compressed from 2465 to 50 using four single layer AEs.

## 3      Results

Only SNPs with p-values lower than $1\times10^{-2}$ were considered for machine learning analysis. First, a SAE is used to learn the deep features of a subset of 2465 SNPs (p-value $< 1\times10^{-2}$) in an unsupervised manner, to capture information about important SNPs and the cumulative epistatic interactions between them. This task is conducted layer wise by stacking four single layer AEs with 2000-1000-500-50 hidden units, where the original 2465 SNPs are compressed into progressively smaller hidden layers (see Fig. 2). The final SAE hidden layer is then used to initialise a deep learning network, based on a multi-layer feedforward ANN (softmax classifier). The classifier is trained with stochastic gradient descent using back-propagation and fine-tuned to classify case-control instances in the validation and test sets. Consequently, four classification experiments were conducted.

The exactness of a classification can be evaluated by computing a contingency table. In this study, classifier performance is assessed through sensitivity (SE), specificity (SP), gini, logloss, area under the curve (AUC) and mean squared error (MSE) as performed in [16], [17]. Classifiers with good predictive capacity possess SE, SP, gini and AUC values close to 1 but logloss and MSE values close to 0.

The data set is split randomly into training (60%), validation (20%) and testing (20%).

### 3.1 Hyperparameters selection.

To maximise the predictive capacity of the classifiers, the network architecture and the regularization parameters were tuned. Adaptive learning rate was used for stochastic gradient descent optimisation, with parameters *rho* and *epsilon* set to 0.99 and $1\times10^{-8}$ respectively, to balance the global and local search efficiencies. Learning rate is configured to 0.005 with rate annealing set to $1\times10^{-6}$ and rate decay set to 1.

More specific tuning parameters were considered for each model in the training phase to obtain optimal results. In the first softmax classifier trained with 2000 compressed hidden units as input, two hidden layers with 10 neurons were used. RectifierWithDropout was used as the activation function throughout the network. To prevent overfitting and to add stability and improve generalisation, Lasso (L1) and Ridge (L2) regularisation values were set to $4.7\times10^{-5}$ and $2.0\times10^{-5}$ respectively. In the second classifier (1000 compressed hidden units as input), two hidden layers with 20 neurons each, a RectifierWithDropout activation function, and $L1=8.5\times10^{-5}$ and $L2=6.0\times10^{-6}$ regularisation values were considered this time for training purposes. The third classifier with 500 compressed hidden units as input, was trained considering two hidden layers with 20 neurons each, a TanhWithDropout activation function, and $L1=6.8\times10^{-5}$ and $L2=5.1\times10^{-5}$ regularisation values. The final classifier considered in this study, with the lowest number of compressed hidden units as input (50 compressed units), was trained with two hidden layers of 50 neurons each, a RectifierWithDropout activation function, and $L1=4.2\times10^{-5}$ and $L2=6.1\times10^{-5}$ regularisation values. Based on empirical analysis, these configurations produced the best results.

### 3.2 Classifier performance

The first layer composed of 2000 compressed hidden units was used to initialize and fine-tune a classifier model. An optimised F1 threshold value of 0.4977 was used to extract the validation set metrics as indicated in Table 1. Successive layers of the SAE were used to initialise and fine-tune the remaining softmax models with 1000, 500 and 50 hidden compressed units as input respectively. On this occasion, metrics were obtained using optimised F1 threshold values 0.6188, 0.4978 and 0.2701 for each of the models respectively.

Table 1. Performance metrics for validation set.

| Compression | SE | SP | Gini | Logloss | AUC | MSE |
|---|---|---|---|---|---|---|
| 2000 | 0.920213 | 0.938326 | 0.960798 | 0.181744 | 0.9804 | 0.054684 |
| 2000, 1000 | 0.840426 | 0.938326 | 0.903365 | 0.288873 | 0.95168 | 0.084837 |
| 2000, 1000, 500 | 0.867021 | 0.889868 | 0.882815 | 0.31456 | 0.94141 | 0.09635 |
| 2000, 1000, 500, 50 | 0.920213 | 0.577093 | 0.697629 | 0.477655 | 0.84881 | 0.159281 |

Table 2 shows the performance metrics obtained using the test set. Optimised F1 threshold values 0.5363, 0.3356, 0.3899 and 0.4615 were used to obtain these metrics by training the models with 2000, 1000, 500 and 50 compressed input units respectively.

Early stopping, particularly helpful to find the optimal number of epochs, was adopted to avoid overfitting. Model building stops when the logloss on the validation set does not improve by at least 1 percent for 2 consecutive scoring epochs (stopping rounds) as shown in Fig. 3. Moreover, in Fig. 3 the AUC plots provide useful information about early divergence between the training and validation curves, highlighting if overfitting occurs.

Table 2. Performance metrics for test set.

| Compression | SE | SP | Gini | Logloss | AUC | MSE |
|---|---|---|---|---|---|---|
| 2000 | 0.949153 | 0.933014 | 0.949936 | 0.1956 | 0.97497 | 0.054057 |
| 2000-1000 | 0.915254 | 0.875598 | 0.910253 | 0.294813 | 0.95513 | 0.087531 |
| 2000-1000-500 | 0.909605 | 0.875598 | 0.900468 | 0.285151 | 0.95023 | 0.087162 |
| 2000-1000-500-50 | 0.785311 | 0.799043 | 0.703566 | 0.476864 | 0.85178 | 0.156315 |

**Model Selection**

The cut-off values for the false and true positive rates in the test set are depicted by the ROC curves in Fig. 4. The ROC curves show a gradual deterioration in the performance of the softmax classifiers as the initial 2465 features (SNPs) are progressively compressed down to 50 hidden units in the SAE. Despite the observable deterioration, the results remain high with 50 compressed hidden units.

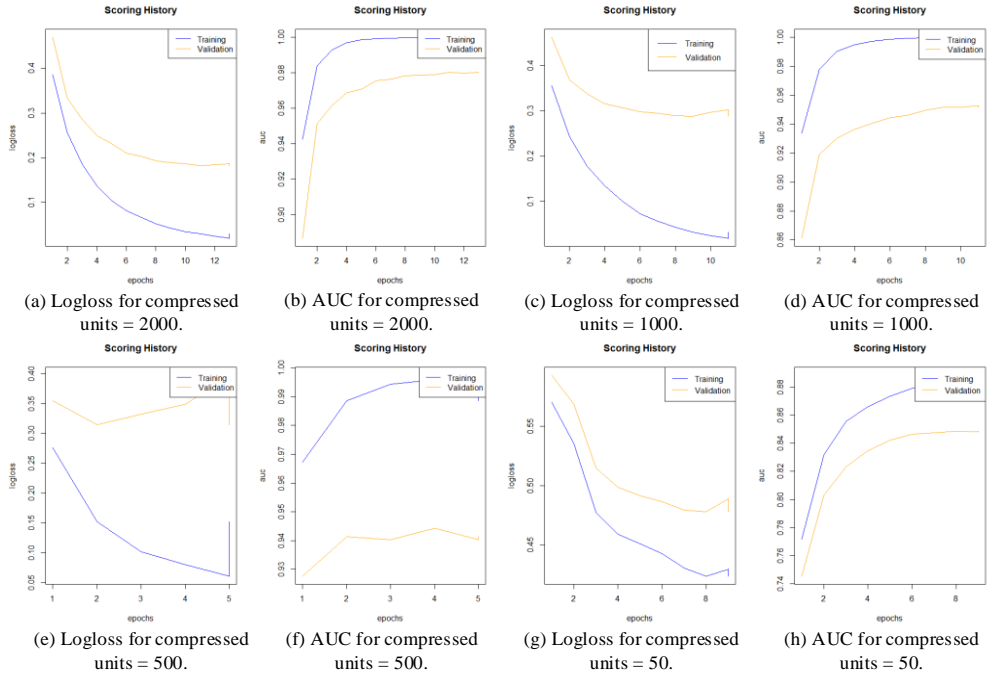

(a) Logloss for compressed units = 2000.
(b) AUC for compressed units = 2000.
(c) Logloss for compressed units = 1000.
(d) AUC for compressed units = 1000.
(e) Logloss for compressed units = 500.
(f) AUC for compressed units = 500.
(g) Logloss for compressed units = 50.
(h) AUC for compressed units = 50.

**Fig. 3.** From (a) to (h), Logloss and AUC plots against epochs for 2000-1000-500-50 compressed units.

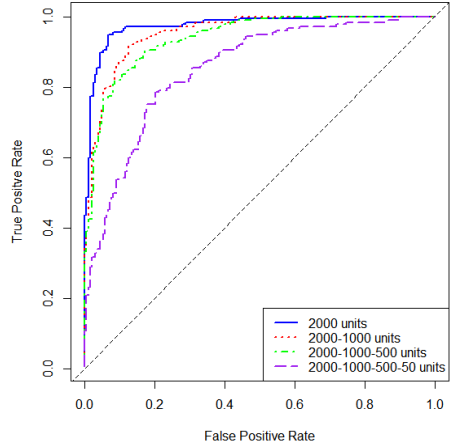

**Fig. 4.** ROC curves for the test set using trained models with the different compressed units.

## 4  Discussion

After QC, 1,997 individuals (879 cases and 1,118 controls) and 240,950 genetic variants remained for subsequent analysis. Logistic regression under an additive genetic model was performed to assess the association between SNPs and binary disease status. However, in logistic regression analysis, SNPs are independently tested for association with the phenotype under investigation, omitting epistatic interactions between these genetic variants. Hence, we performed unsupervised feature extraction in a set of 2465 SNPs (p-value $< 1\times10^{-2}$) stacking four single layer AEs with 2000-1000-500-50 hidden units. To evaluate the effectiveness of our SAE extracted features, we used a supervised learning model to classify extremely obese samples from normal control samples. Four multi-layer feedforward ANN (softmax) were trained with the compressed hidden units considered in the SAE.

Overall, performance metrics obtained using the test set showed slight better improvement in classification accuracy than metrics obtained in the validation set. The results were only higher in the validation set when 2000 features were used as can be observed in Table 1. The best result using the test set was obtained using 2000 features (SE=0.949153, SP=0.933014, Gini=0.949936, Logloss=0.1956, AUC=0.97497 and MSE=0.054057). Conversely, the lowest performance in the test set was achieved when the features were compressed to 50 hidden units (SE=0.785311, SP=0.799043, Gini=0.703566, Logloss=0.476864, AUC=0.85178 and MSE=0.156315). In fact, performance metrics started deteriorating when the layers of the SAE were gradually compressed (See Table 1 and Table 2). It can be noted that the specificity in the validation set (SP=0.577093) when 50 features were used was noticeable lower than the value obtained with the same number of compressed neurons in the test set (SP=0.799043).

Fig. 3 shows that there is not significant indications of overfitting between the training and validation datasets. On the other hand, Fig. 4 shows a gradual deterioration in performance when the features are compressed into smaller number of hidden units although the performance is still high even with 50 units. Although the results were higher in the classifiers with the higher number of hidden units and less stacked layers, we managed to compress the initial 2456 SNPs to 50 hidden units and still get over 85 % AUC with relatively low overfitting as shown in Fig. 3. This demonstrates the potential of using our deep learning methodology to abstract large, complex and unstructured data into latent representations capable of capturing the epistatic effect between SNPs in GWAS.

Using deep learning stacked autoencoders to initialise the multi-layer feedforward ANN classifier outperformed the results obtained in our previous study [17] using the same dataset. In [17], SNPs were filtered considering suggestive and Bonferroni levels of significance and then used for classification analysis using a multi-layer feedforward ANN. These SNPs were selected based on highly conservative threshold designed to mitigate type I errors but, in our experiments, we demonstrated that significant SNPs had insufficient discriminative capacity to discern between obese and non-obese individuals. Hence, assessing the impact of single variants on the phenotype under study using traditional statistic is insufficient, since epistatic interactions between SNPs are omitted. Our results in this study show an improvement specially in the performance metrics Gini, Logloss and MSE (more reliable models) in comparison with the values obtained in our previous study. Utilising deep learning SAE is a better alternative than using statistical approaches such as logistic regression in GWAS for classification tasks, with potential to help explaining the missing heritability phenomena in polygenic obesity.

This paper presents a novel approach with emphasis in the feature extraction and classification phases, using latent information extracted from high-dimensional genomic data for the identification of individuals with higher predisposition to obesity. However, compressing the features using SAEs makes very difficult to identify what information from the 2465 SNPs contributed to the compress hidden units, mainly due to the lack of interpretation of deep learning models which act as a black box [18]. This limitation fosters the need to create robust methods for the interpretation of deep learning networks.

## 5 Conclusion

In this paper, a novel approach to investigate complex interactions between genetic variants in polygenic obesity, utilising cases and controls from Geisinger MyCode project has been presented. We combined common genetic tools and techniques for QC and association analysis with deep learning to capture relevant information and the epistatic interactions between SNPs. Quality control, association analysis, multi-layer feedforward ANN classifier and deep learning SAE constitute the components of our proposed methodology.

Although the utilization of deep learning SAE and multilayer feedforward ANN (softmax) has been previously considered in many areas of research, we claim that this is the first time that it has been applied to study epistatic interactions between SNPs in GWAS of polygenic obesity. This framework has been successfully employed for pre-term birth classification in African-American woman [16].

Despite we have presented encouraging results, the study needs further research. Mapping SNPs inputs to hidden nodes in the SAE is still a recognized limitation as deep learning approaches act as black box where models become difficult to interpret. In future work, we aim to create a robust method that helps interpreting the outcome of the compressed units by combining association rule mining (ARM) [19] and the strength of SAEs.

# References

[1] W. P. T. James, "WHO recognition of the global obesity epidemic.," *Int. J. Obes. (Lond).*, vol. 32 Suppl 7, pp. S120–S126, 2008.

[2] L. N. Borrell and L. Samuel, "Body Mass Index Categories and Mortality Risk in US Adults: The Effect of Overweight and Obesity on Advancing Death," *Am. J. Public Health*, vol. 104, no. 3, pp. 512–519, Mar. 2014.

[3] A. J. Walley, A. I. F. Blakemore, and P. Froguel, "Genetics of obesity and the prediction of risk for health," *Hum. Mol. Genet.*, vol. 15, no. SUPPL. 2, pp. 124–130, 2006.

[4] K. R. Rao, N. Lal, and N. V Giridharan, "Genetic & epigenetic approach to human obesity," *Indian J. Med. Res.*, vol. 140, no. November 2014, pp. 589–603, 2015.

[5] J. M. Walker, *Genetic Variation*, vol. 628. Totowa, NJ: Humana Press, 2010.

[6] A. E. Locke *et al.*, "Genetic studies of body mass index yield new insights for obesity biology," *Nature*, vol. 518, no. 7538, pp. 197–206, Feb. 2015.

[7] J. H. Moore and S. M. Williams, "Epistasis and Its Implications for Personal Genetics," *Am. J. Hum. Genet.*, vol. 85, no. 3, pp. 309–320, Sep. 2009.

[8] K. Chen and K. Janz, "Re-defining the roles of sensors in objective physical activity monitoring," *Med. Sci. ...*, vol. 44, no. 301, pp. 1–18, 2012.

[9] M. I. McCarthy *et al.*, "Genome-wide association studies for complex traits: consensus, uncertainty and challenges," *Nat. Rev. Genet.*, vol. 9, no. 5, pp. 356–369, May 2008.

[10] W. Li, "Three lectures on case control genetic association analysis," *Brief. Bioinform.*, vol. 9, no. 1, pp. 1–13, Oct. 2007.

[11] S. Purcell *et al.*, "PLINK: A Tool Set for Whole-Genome Association and Population-Based Linkage Analyses," *Am. J. Hum. Genet.*, vol. 81, no. 3, pp. 559–575, Sep. 2007.

[12] G. M. Clarke *et al.*, "Basic statistical analysis in genetic case-control studies," *Nat. Protoc.*, vol. 6, no. 2, pp. 121–133, Feb. 2011.

[13] A. Ng, "Sparse Autoencoder," in *CS294A Lecture notes*, 2011, pp. 1–19.

[14] D. E. Rumelhart, G. E. Hinton, and R. J. Williams, "Learning representations by back-propagating errors," *Nature*, vol. 323, no. 6088, pp. 533–536, Oct. 1986.

[15] Q. V Le, "A Tutorial on Deep Learning Part 2: Autoencoders, Convolutional Neural Networks and Recurrent Neural Networks," Mountain View, CA, 2015.

[16] P. Fergus, C. A. Curbelo, B. Abdulaimma, P. Lisboa, and C. Chalmers, "Utilising Deep Learning and Genome Wide Association Studies for Epistatic-Driven Preterm Birth Classification in African-American Women," *arXiv:1801.02977*, pp. 1–11, Jan. 2018.

[17] C. A. Curbelo, P. Fergus, A. C. Montañez, A. Hussain, D. Al-Jumeily, and C. Chalmers, "Deep Learning Classification of Polygenic Obesity using Genome Wide Association Study SNPs," *arXiv:1804.03198*, p. 8, Apr. 2018.

[18] Z. C. Lipton, "The Mythos of Model Interpretability," *arXiv:1606.03490*, Jun. 2016.

[19] R. Agrawal, T. Imieliński, and A. Swami, "Mining association rules between sets of items in large databases," *Proc. 1993 ACM SIGMOD Int. Conf. Manag. Data*, vol. 22, no. 2, pp. 207–216, Jun. 1993.